\documentclass[twocolumn,showpacs,preprintnumbers,amsmath,amssymb,superscriptaddress]{revtex4}

\usepackage{graphicx}
\usepackage{dcolumn}
\usepackage{bm}

\begin{document}


\title{Single carbon nanotube transistor at GHz frequency}

\author{J. Chaste}
\affiliation{Ecole Normale Sup\'erieure, Laboratoire Pierre Aigrain,
24 rue Lhomond, 75005 Paris, France } \affiliation{CNRS UMR8551,
Laboratoire associ\'e aux universit\'es Pierre et Marie Curie et
Denis Diderot, France }
\author{L. Lechner}
\affiliation{Low Temperature Laboratory, Helsinki University of
Technology, Otakaari 3A, Espoo, 02015 Finland}
\author{P. Morfin}
\affiliation{Ecole Normale Sup\'erieure, Laboratoire Pierre Aigrain,
24 rue Lhomond, 75005 Paris, France } \affiliation{CNRS UMR8551,
Laboratoire associ\'e aux universit\'es Pierre et Marie
Curie et Denis Diderot, France }
   \author{G. F\`eve}
 \affiliation{Ecole Normale Sup\'erieure, Laboratoire Pierre
Aigrain,  24 rue Lhomond, 75005 Paris, France } \affiliation{CNRS
UMR8551,  Laboratoire associ\'e aux universit\'es Pierre et Marie
Curie et Denis Diderot, France }
\author{T. Kontos}
\affiliation{Ecole Normale Sup\'erieure, Laboratoire Pierre Aigrain,
24 rue Lhomond, 75005 Paris, France } \affiliation{CNRS UMR8551,
Laboratoire associ\'e aux universit\'es Pierre et Marie
Curie et Denis Diderot, France }
  \author{J.-M. Berroir}
\affiliation{Ecole Normale Sup\'erieure, Laboratoire Pierre Aigrain,
24 rue Lhomond, 75005 Paris, France } \affiliation{CNRS UMR8551,
Laboratoire associ\'e aux universit\'es Pierre et Marie
Curie et Denis Diderot, France }
\author{D.C. Glattli}
\affiliation{Ecole Normale Sup\'erieure, Laboratoire Pierre Aigrain,
24 rue Lhomond, 75005 Paris, France } \affiliation{CNRS UMR8551,
Laboratoire associ\'e aux universit\'es Pierre et Marie Curie et
Denis Diderot, France }\affiliation{Service de Physique de l'Etat
Condens\'{e}, CEA Saclay, F-91191 Gif-sur-Yvette, France }
\author{H. Happy}
 \affiliation{Institut d'Electronique, de Micro\'electronique et de Nanotechnologie,
 UMR-CNRS 8520, BP 60069, Avenue Poincar\'e, 59652, Villeneuve d'Asq, France }
 \author{P. Hakonen}
\affiliation{Low Temperature Laboratory, Helsinki University of
Technology, Otakaari 3A, Espoo, 02015 Finland}
\author{B. Pla\c{c}ais}
   \email{Bernard.Placais@lpa.ens.fr}
\affiliation{Ecole Normale Sup\'erieure, Laboratoire Pierre Aigrain,
24 rue Lhomond, 75005 Paris, France } \affiliation{CNRS UMR8551,
Laboratoire associ\'e aux universit\'es Pierre et Marie
Curie et Denis Diderot, France }

\date{\today}

\begin{abstract}
We report on microwave operation of top-gated single carbon nanotube
transistors. From transmission measurements in the
$0.1$--$1.6\;\mathrm{GHz}$  range we deduce device transconductance
$g_m$ and gate-nanotube capacitance $C_g$ of micro- and nanometric
devices. A large and frequency-independent $g_m\sim 20 \;\mathrm{\mu
S}$ is observed on short devices which meets best dc results. The
capacitance per unit gate length $\sim 60\; \mathrm{aF/\mu m}$ is
typical of top gates on conventional oxide with $\epsilon\sim10$.
This value is a factor $3$--$5$ below the nanotube quantum
capacitance which, according to recent simulations, favors high
transit frequencies $f_T=g_m/2\pi C_{g}$. For our smallest devices,
we find a large $f_T\sim 50\; \mathrm{GHz}$ with no evidence of
saturation in length dependence.


 \end{abstract}

\maketitle

Carbon nanotube field effect transistors (CNT-FETs) are very
attractive as ultimate, quantum limited devices. In particular,
ballistic transistors have been predicted to operate in the the
sub-THz range \cite{Javey2003,Burke2004}. Experimentally, a
state-of-the-art cut-off frequency of $30 \;\mathrm{GHz}$ has been
reached in a low impedance multi-nanotube device
\cite{LeLouarnAPL2007}, whereas $8\;\mathrm{GHz}$ was achieved with
a multigate single nanotube transistor \cite{Wang2007IEEE}. Indirect
evidence of microwave operation were also obtained in experiments
based on mixing effects or channel conductance measurement in single
nanotubes
\cite{Appenzeller2004APL,Yu2006APL,Li2004NanoLetters,Rosenblatt2005APL,Pesetski2006APL}.

The extraordinary performances of nanotubes as molecular field
effect transistors rely on a series of unique properties.
High-mobility "p-doped" single walled nanotubes can be obtained
 by CVD-growth, with  a semiconducting gap
$\Delta\sim 0.5$--$1\; \mathrm{eV}$ (diameter 1-2 nm)
\cite{Dufkop2004Nanoletters}. Low Schottky-barrier contacts, with Pd
metallisation, and quasi-ballistic transport result in a channel
resistance $R_{ds}$ approaching the quantum limit, $h/4e^2=6.5\;
\mathrm{k\Omega}$ for a 4-mode single walled nanotube
\cite{Hermann2007PRL}. High saturation currents, limited by optical
phonon emission, allow large biases, $I_{ds}\sim 20\; \mathrm{\mu
A}$ at $V_{ds}\gtrsim 1\;\mathrm{V}$ in short nanotubes
\cite{Yao2000PRL,Bourlon2004PRL}. The above numbers and the good
gate coupling  explain the large transconductances, $g_m\sim
I_{ds}/\Delta\gtrsim 10\; \mathrm{\mu S}$, observed in dc
experiments \cite{Javey2003}. In the ac, an intrinsic limitation is
given by the transit frequency $f_T=g_m/2\pi C_{g}$, where $C_g$ is
the gate-nanotube capacitance. Here $C_g=C_{geo}C_Q/(C_{geo}+C_Q)$
is the series combination of the quantum and geometrical
capacitances, $C_Q$ and $C_{geo}$. Ultrathin oxide coating in
CNT-FETs allows to approach the quantum limit with a capacitance per
unit gate length $l_g$,
 $C_{geo}/l_g>C_Q/l_g=4e^2/hv_F \sim 400\; \mathrm{aF/\mu m}$ for
  $v_F\sim 4\times 10^5 \; \mathrm{m/s}$, a typical value for semiconducting NTs \cite{Retailleau2007}.

\begin{figure}
\includegraphics[scale=0.5]{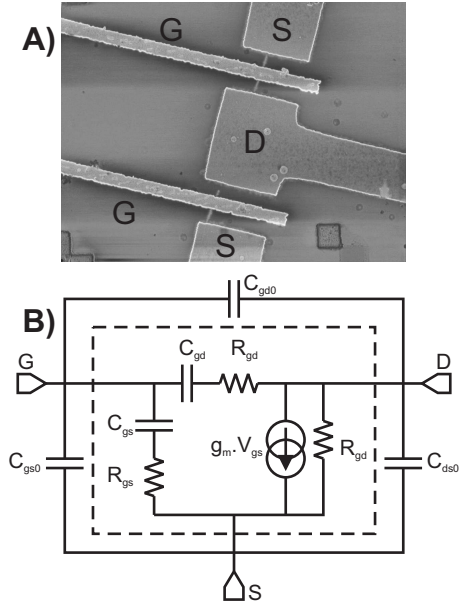}
\caption{Panel A : scanning electron microscope image of a single
carbon-nanotube double-gate transistor. Gate length is $300
\;\mathrm{nm}$. Panel B :  small-signal equivalent circuit of the
carbon-nanotube transistor. Extrinsic components are displayed
outside the dashed rectangle. Notations are explained in the
text.}\label{sample_image}
\end{figure}

Beside basic interest for quantum limited nano-devices, single
nanotube transistors offer new opportunities for fast charge
detection due to unique combination of short time response and high
charge sensitivity. At present, charge counting experiments
performed in nanotube or semiconducting quantum dots use either
single electron transistors \cite{Biercuk2006,Roschier2004} or
quantum point contact detectors \cite{Gustavsson2006} which operate
on microsecond timescales. High sensitivity of NT-FET has been
demonstrated recently by monitoring tunneling events between the
nanotube and a nearby gold particle \cite{Gruneis2007} at the dc
limit. Performing such experiments with quantum dots at nanosecond
timescales would allow to extend charge counting in the coherent
regime relevant for full quantum electronics \cite{Feve2007}. Given
the shot noise limitation of a nanotube detector of bandwidth
$1/\tau=1 \mathrm{ns^{-1}}$, overestimated by the poissonian value
$\sqrt{I_{ds}\tau/e}= 250\;\mathrm{electrons}$ for $I_{ds}\sim 10\;
\mathrm{\mu A}$, single charge resolution requires a charge gain
$g_m\tau/C_{g}\gtrsim 250$ or equivalently a transit frequency
$f_T=g_m/2\pi C_{g}\gtrsim 40\;\mathrm{GHz}$.

In this letter we demonstrate room-temperature broad-band GHz
operation ($f=0.1$--$1.6\,\mathrm{GHz}$) of nano-transistors made of
a single carbon nanotube, with  $l_g= 0.3\;\mathrm{\mu m} $ and $ 3
\;\mathrm{\mu m}$ (channel lengths $1$ and $ 3.5\;\mathrm{\mu m}$
respectively). Our main results are the high $g_m\sim
10$--$20\;\mathrm{\mu S}$ in short tubes, and the scaling
$C_{g}\propto l_g$. From the latter we deduce $C_{g}/l_g\sim
60\;\mathrm{aF/\mu m} $, in accordance with previous low-frequency
determinations \cite{Ilani06NaturePhysics}, and $f_T\sim
50\;\mathrm{GHz}$ for our $l_g=300\;\mathrm{nm}$ devices.

\begin{figure}[ttt]
\includegraphics[scale=1.1]{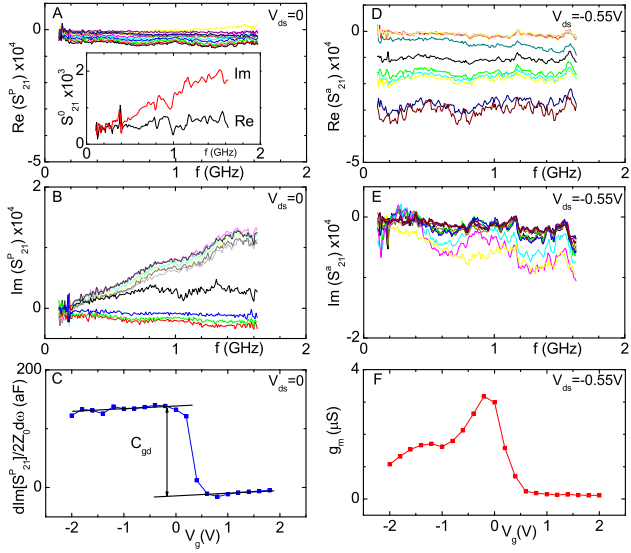}
\caption{ A $3\;\mathrm{\mu m}$ gate transistor. Transmission
amplitude $S_{21}$ as function of frequency. Background contribution
(inset of panel A), passive zero-bias (panels A and B) and active
biased (panels D,E) contributions. Panels A and D (resp. B and E)
have same scales for direct data comparison. Lines correspond to
$V_g=-2$--$2\mathrm{V}$ with an increment of $0.4\mathrm{V}$. Panel
C gives the slope $\langle d\Im[S^p_{21}(\omega)]/d\omega\rangle$ as
function of gate voltage. It shows a discontinuity at the opening of
channel conductance associated with the gate-drain capacitance.
Panel F shows device transconductance. } \label{XL_rf}
\end{figure}

Double-gate carbon nanotube transistors were fabricated by e-beam
lithography in a coplanar strip-line geometry, with two symmetric
top gates (see Fig.\ref{sample_image}), on oxidized high-resistivity
silicon substrates (resistivity $3$--$5 \; \mathrm{k\Omega.cm}$).
Nanotubes were synthetized from nano patterned catalyst pads using a
standard CVD recipe.
Palladium contact evaporation is followed by multistep oxidation of
thin aluminium, for a total oxide thickness of $\simeq 6 \;
\mathrm{nm}$, and finally gold-gate deposition. This process
provides full oxidation of Al into Al$_2$O$_3$  with $\epsilon\simeq
9.8$. From sample geometry and dielectric constants, we estimate
$C_{geo}/l_g\sim 100\; \mathrm{aF/\mu m}$.

Devices were characterized in a standard $S$-parameter measurement
using an  RF probe station and a network analyzer. Attenuators
($-6\;\mathrm{dB}$ and $-10\;\mathrm{dB}$), followed by bias tees,
are mounted directly at the input and output ports of the station to
minimize standing waves in the $Z_0=50\; \mathrm{\Omega}$ cables due
to large impedance mismatch $R_{ds}/Z_0\sim300$. To compensate for
small voltage gain, $Z_0g_m\sim 10^{-3}$, output signals are
amplified by two $0.1$--$2\; \mathrm{GHz}$ bandwidth low noise
amplifiers in series. As a consequence, $S$-parameter measurement is
restricted to the transmission coefficient $S_{21}(V_{ds},V_{g})$,
which in turn can be accurately calibrated.

\begin{figure}[ttt]
\includegraphics[scale=1.4]{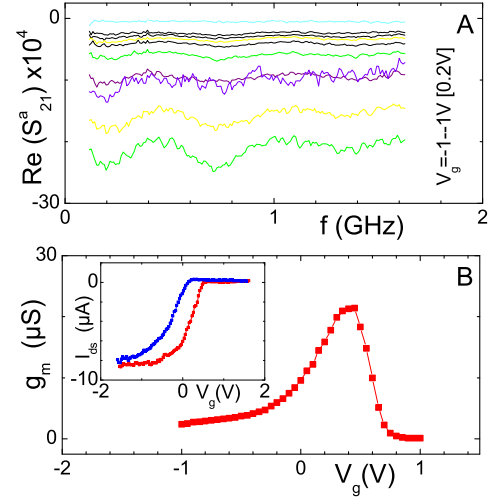}
\caption{ Characterisation of a  $300\;\mathrm{nm}$ gate transistor.
Transmission amplitude of active contribution (panel A) and
transconductance (panel B). Inset shows hysteresis in the dc
characteristics as $V_g$ is swept. }\label{sample_M_AB43}
\end{figure}

Data reduction and analysis are based on the standard equivalent
circuit of Fig.\ref{sample_image}-B. Extrinsic elements,
$C_{gs0}\sim C_{ds0}\sim 10$--$20\; \mathrm{fF}$, estimated from
independent reflection measurements, and $C_{gd0}\sim 2\;
\mathrm{fF}$ measured as explained below, are minimized by the
coplanar sample design. The gate capacitance $C_g=C_{gd}+C_{gs}$ is
splitted into gate-drain $C_{gd}$ and gate-source $C_{gs}$
contributions. These are associated to charge relaxation resistances
$R_{gd}$ and $R_{gs}$. We have $R_{gd}\gg 1/C_{gd}\omega$  in the
OFF state ($V_g\gtrsim1\;\mathrm{V}$) and $R_{gd}\ll 1/C_{gd}\omega$
in ON state ($V_g\lesssim0\;\mathrm{V}$). Considering that
$Z_0C_{gs0}\omega\sim 10^{-2}$, $Z_0C_{gd0}\omega\sim 10^{-3}$ and
$Z_0g_m\sim 10^{-3}$ in the low-frequency range $\omega/2\pi\leq 1.6
\; \mathrm{GHz}$ of our experiment, we obtain
\begin{eqnarray}
S_{21}&=&S^0_{21}+S^p_{21}(V_{g})+S^a_{21}(V_{ds},V_{g})\quad
\mathrm{with}\nonumber \\
S^0_{21}&\approx &\jmath \;2Z_0\omega C_{gd0}  \label{dummy}\\
 S^p_{21}&\approx & \jmath\; 2Z_0\omega C_{gd}/(1+\jmath\omega C_{gd}R_{gd})   \label{passive}\\
 S^a_{21}&\approx &-2Z_0g_m +  \Delta S^p_{21}(V_{ds})  \label{active}
\label{simplified}\end{eqnarray} being, respectively, the background
transmission amplitude in the pinched state, the passive
contribution at the onset of channel conduction in the zero bias
state, and the active contribution under bias. Note that $\Delta
S^p_{21}=S^p_{21}(V_{ds})-S^p_{21}(0)$ constitutes the imaginary
part of $S^a_{21}$.

In order to check the validity of our analysis, we first consider
the case of long FETs ($l_{g}=3\; \mathrm{\mu m}$) where gate
capacitance contributions are easily resolved. Typical datas are
exhibited in Figs.\ref{XL_rf}. The inset of Fig.\ref{XL_rf}.A shows
both quadratures of the background transmission $S^0_{21}(f)$. From
the slope of $\Im(S^0_{21})(f)$ and Eq.(\ref{dummy}) we deduce the
small $C_{gd0}\sim 1.5 \; \mathrm{fF}$. Substraction of this
contribution, keeping zero bias conditions, gives the
$V_g$-dependent passive contribution $S^p_{21}(f)$ shown in
(Figs.\ref{XL_rf}.A,B). In the ON state $\Im(S^p_{21})$ has a linear
frequency dependence, whose slope is plotted in Fig.\ref{XL_rf}-C as
a function of $V_g$. The pinch-off transition appears here as a
strong discontinuity associated with the divergence of the charge
relaxation resistance $R_{gd}$. From the step amplitude and using
Eq.(\ref{passive}), we deduce $C_{gd}\simeq 170\;\mathrm{aF}$. This
value is very representative of the five long samples which we have
measured with an average of $160\;\mathrm{aF}$ and a standard
deviation of $50\;\mathrm{aF}$. A small monotonic dependence in
$\Re(S^p_{21})(V_g)$ (Fig.\ref{XL_rf}.A) is also observed which is
possibly due to a residual substrate conduction, not taken into
account in our analysis.

The active contribution (Figs.\ref{XL_rf}.D-E) is obtained from the
bias dependence of $S_{21}(f)$ at constant $V_g$. The in-phase
signal $\Re(S^a_{21})(f)$ is prominent and almost frequency
independent. Slow frequency oscillations are reminiscent of
calibration imperfections. Averaging $\Re(S^a_{21})$ over
$0.2$--$1.6\; \mathrm{GHz}$ gives, according to Eq.(\ref{active}),
the transconductance as function of $V_g$ shown in
Fig.\ref{XL_rf}.F. It has a maximum $g_m\lesssim 3\;  \mathrm{\mu
S}$ close to the maximum observed at the dc ON-OFF transition.
Representative values are $g_m\lesssim 1$--$4\;  \mathrm{\mu S}$.
The small negative part, $\Im(S^a_{21})\sim-\Im(S^p_{21})$ in
Fig.\ref{XL_rf}.E, is a capacitive contribution due to a shift of
the pinch-off under bias.
Altogether, these measurements show that one can quantitatively
analyze the dynamical properties of a single nanotube transistor.
With $g_m\sim1.5 \mathrm{\mu S}$ and $C_g\sim85\mathrm{aF}$ (per
gate finger), the transit frequency of the $3\; \mathrm{\mu m}$
device of Fig.\ref{XL_rf} is $f_T\sim 3\;\mathrm{GHz}$.

For short NT-FETs ($l_{g}=300\; \mathrm{nm}$ in
Fig.\ref{sample_image}) we observe smaller $C_{gd}$ and larger $g_m$
due to smaller channel resistance. Example shown in
Fig.\ref{sample_M_AB43} represents a new state-of-the-art for RF
operation with a maximum of $g_m(V_g)\gtrsim 20\;\mathrm{\mu S}$
(double gate fingers). Gate drain capacitance is $C_{gd}\simeq
35\;\mathrm{aF}$ (standard deviation $25\;\mathrm{aF}$). Average
values for the six short NT-FETs measured are $g_m\simeq
12\;\mathrm{\mu S}$ and $C_{gd}\simeq 35\;\mathrm{aF}$ (double gate
fingers). As seen on the dc characteristics in the inset of
Figs.\ref{sample_M_AB43}.B, hysteresis is observed in these
measurements which has been omitted elsewhere for clarity. Comparing
our data for the $300\,\mathrm{nm}$ and $3\;\mathrm{\mu m}$ devices
shows that, within experimental uncertainty, $C_{gd}$ scales with
$l_g$. Assuming symmetric distribution of gate-drain and gate-source
capacitances at zero bias, we obtain $C_g/l_g\simeq
2C_{gd}/l_g\simeq 60\;\mathrm{aF/\mu m}$. With $g_m\sim10
\mathrm{\mu S}$ and $C_g<30\mathrm{aF}$ (per gate finger), the
transit frequency of the $300\; \mathrm{nm}$ device of
Fig.\ref{sample_M_AB43} is $f_T\gtrsim 50\;\mathrm{GHz}$.

These measurements show that NT-FET performances do improve
drastically upon gate size reduction down to the nanometric scale
(factor of 15 in $f_T$ for a factor 10 in size). We now discuss
possible routes for improvements. First one is to increase $g_m$ by
reducing access resistance due to the ungated NT sections at drain
and source (see Fig.\ref{sample_image}-A). This effect is
significant on devices using deposited CNTs, it is minimized with
\emph{in situ} CVD-grown NTs which are naturally p-doped in the
absence of gating. Actually ungated regions are needed to reduce
direct gate-drain/source capacitive coupling. Our values for gate
capacitance are close to estimate for the geometrical contribution
and in accordance with previous low-frequency measurements in
similar top-gated devices \cite{Ilani06NaturePhysics}. They are
still smaller than above numbers for the quantum capacitance. One
may wonder whether better performance could be achieved by working
closer to the quantum limit in using high-$\kappa$ and/or thinner
oxides. This would improve gate coupling, but at the same time
slow-down electron dynamics due to the screening of electronic
interactions as discussed in Ref.\cite{Gabelli07PRL}.  In the
absence experimental data and theoretical model, we shall rely on
recent numerical simulations showing that transit frequency is
maximized for $C_Q\gtrsim 3$--$5\times C_{geo}$
\cite{Retailleau2007}. This condition which is close to our
experimental realization. The most promising route is further size
reduction below 100 nm as we have not observed evidence for
saturation down to 300nm. Finally noise performance remain to be
characterized, in particular the conditions for shot-noise limited
resolution.

In conclusion, we have demonstrated high-transconductance SWNT-FET
properties up to $1.6\; \mathrm{GHz}$. We observe that high
sensitivity is preserved, and that gate capacitance scales with gate
length down to $300\; \mathrm{nm}$. Transit frequencies as high as
$50\;\mathrm{GHz}$ have been inferred, indicating that nanotube FETs
are promising fast sensors.

\begin{acknowledgments}

Authors acknowledge fruitful discussions with S. Galdin-Retailleau,
P. Dollphus, J.P. Bourgoin, V. Derycke, and G. Dambrine. The
research has been supported by french ANR under contracts
ANR-2005-055-HF-CNT, ANR-05-NANO-010-01-NL-SBPC, and EC STREP
project CARDEQ under contract IST-FP6-011285.

\end{acknowledgments}

\end{document}